# A Matlab toolbox for fractional relaxation-oscillation equations

Song Wei, Wen Chen[*]

*Institute of Soft Matter Mechanics, Department of Engineering Mechanics, Hohai University,*

*Nanjing 210098, P.R. China*

## Abstract

Stress relaxation and oscillation damping of complex viscoelastic media often manifest history- and path-dependent physical behaviors and cannot accurately be described by the classical models. Recent research found that fractional derivative models can characterize such complex relaxation and damping. However, to our best knowledge, easy-to-use numerical software is not available for fractional relaxation-oscillation (FRO) equations. This paper is to introduce an open source free Matlab toolbox which we developed in recent years for numerical solution of the FRO equations. This FRO toolbox uses the predictor-corrector approach for the discretization of time fractional derivative, and non-expert users can accurately solve fractional relaxation-oscillation equations via a friendly graphical user interface. Compared with experimental data, our numerical experiments show that the FRO toolbox is highly efficient and accurate to simulate viscoelastic stress relaxation and damped vibration. This free toolbox will help promote the research and practical use of fractional relaxation-oscillation equations.

## 1. Introduction

Stress relaxation and damped vibration are common in nature and engineering. Stress relaxation plays a key role in the performance of many engineering components, while damped vibration is often encountered in quake-proof structures and noise-reducing devises. In history, Maxwell and Debye initiated research of relaxation and damping, and Zener [1] proposed an exponential model to describe the relaxation of the strain field by considering the connection between diffusion and relaxation. In 1921, Nutting developed a stress-strain relation underlying an inverse power-law relaxation [2]. In 1970, stretched exponential was presented by G. Williams and D.C. Watts [3] to describe dielectric spectra of polymers. These classical models have found successful applications in many areas of sciences and engineering.

[*] Email: chenwen@hhu.edu.cn

However, neither the Nutting law nor the Kohlrausch-Williams-Watts relaxation law is compatible with experimental data over the full range [4]. It is also widely observed that relaxation and damping of complex viscoelastic media characterize history-dependent features and cannot be accurately described by the classic differential equation models of integer order [5]. As a result, such behaviors are usually called complex or anomalous in literature. It is worth noting that although these classical differential equation models, regardless of linearity or non-linearity, succeed in describing some particular cases of the above-mentioned anomalous relaxation and damping, some contrived parameters are often required which do not have clear physical significance only to fit the experiment data.

Based on the standard Zenner model, Glöckle and Nonnenmacher [6] made a breakthrough to propose a well-posed fractional Zenner model, in which the integer-order derivative term of relaxation and damping in the classical models are replaced by the fractional derivative term, e.g., the fractional Maxwell model. Mainardi [7] also pioneered researches on fractional derivative modeling of complex relaxation-oscillation behaviors. Furthermore, Friedrich [8] established the connection between the fractional derivative constitutive equations and molecular dynamic theory. Metzler and Nonnenmacher [1] investigate anomalous diffusion with stress relaxation through fractional derivative model. More recently, Chen et al. [9] developed novel positive fractional derivative and fractal derivative models of complex relaxation-oscillation.

Recent researches show that the fractional derivative models can characterize history-dependent relaxation and damping behaviors and require fewer yet easier-to-get parameters than the traditional models of integer order derivative. On the other hand, it is difficult to get the analytic solution of most fractional derivative equations (FDEs). Even if analytic solution can be obtained, it is difficult to utilize due to its complex form. Therefore, numerical solution is the applicable way to model a wide range of practical problems. It is also known that the fractional derivative is an integro-differential operator and its numerical solution is more challenging than the differential equation of integer order. Nowadays one can find few toolboxes for numerical solution of fractional derivative equations such as CRONE [10], FOMCON [11]. However, to our best knowledge, easy-to-use numerical software is not available for fractional relaxation-oscillation (FRO) equations. This paper is to introduce an open source free Matlab toolbox with the purpose to fill this gap which we developed in recent years for numerical solution of the FRO equations. Non-expert users can use this toolbox to accurately solve fractional relaxation-oscillation equations via a friendly graphical user interface.

Various numerical methods for solving fractional derivative equations have been developed in recent decades, such as Adomian decomposition method [12, 13], generalized Taylor matrix method [14], collocation method [15], variational iteration method [16], and operational matrix method [17]. We consider the predictor-corrector method [18, 19, 20] is a simple and powerful technique to solve the FRO equation and is used in our FRO toolbox. The codes of this FRO toolbox is free to download and further modify.

The rest of this paper is organized as follows. Section 2 introduces the fractional relaxation-oscillation equations and the numerical algorithm. Section 3 describes our FRO toolbox, followed by Section 4 where several numerical examples are presented to illustrate the implementation and use of the FRO toolbox. Finally, Section 5 concludes this paper with some remarks.

## 2. Predictor-corrector approach for fractional relaxation-oscillation equations

Since the 1980s, the fractional order calculus and its applications have widely been concerned [21], involving bioengineering [22, 23], random variable in engineering [24], viscoelastic dampers [25, 26, 27], control [28], complex media [29, 30, 31], unconventional statistics [32], just to mention a few. Up to now, there are various definitions of fractional calculus, among which the often used ones include Riemann-Liouville definition, Grünwald-Letnikov definition and Caputo definition. Compared with Riemann-Liouville fractional derivative, Caputo fractional derivative does not include fractional-order initial or boundary conditions and its Laplace transform is easy-to-get. Thus, the Caputo definition is often employed to model physical and mechanical problems and is used in our FRO toolbox.

### 2.1. Fractional relaxation-oscillation equation

Lab experiments and field observations indicate that many viscoelastic materials appear to have "memory" underlying a mechanics property between elastomer and viscous fluid. It is known that the constitutive relations of elasticity and viscous fluid are $\sigma(t) \sim d^0 \varepsilon(t)/dt^0$ and $\sigma(t) \sim d^1 \varepsilon(t)/dt^1$, respectively. From the fractional derivative perspective, one can intuitively guess that viscoelastic media may have a constitutive relation $\sigma(t) \sim d^\alpha \varepsilon(t)/dt^\alpha$ ( $0 < \alpha < 1$ ) in between elastomer and viscous fluid [21, 33]. This happens to be true and the fractional model is verified in many research reports. Based on this constitutive relationship, the fractional relaxation-oscillation equation is a successful phenomenological model to describe history-dependent behaviors. Compared with the classical integer-order equation models, the fractional relaxation equations have advantage in the description of superslow processes, while fractional oscillation equations found an edge to depict intermediate processes [7]. With spatial discretization taken into account, relaxation and damping could be considered a simple process of diffusion and fluctuation,

respectively. The fractional relaxation-oscillation equation is given by

$$\begin{cases} {}_0^C D_t^\alpha u(t) + Au(t) = f(t) \\ y^{(k)}(0) = y_0^{(k)}, (k = 0,1\cdots, n-1) \end{cases} \quad (1)$$

When $0 < \alpha \leq 1$, Eq. (1) depicts a relaxation process underlying power law attenuation [34]. When $1 < \alpha \leq 2$, Eq. (1) describes a oscillation process which are highly related to the fractional order $\alpha$ and "natural frequency" $\omega_0$ [34]. In Eq. (1), $A$ represents relaxation coefficient, and Caputo fractional derivative ${}_0^C D_t^\alpha f(t)$ is defined by [21]

$$ {}_0^C D_t^\alpha f(t) = \frac{1}{\Gamma(\alpha - n)} \int_a^t \frac{f^{(n)}(\tau)}{(t-\tau)^{\alpha+1-n}} d\tau \quad (n = \lfloor \alpha \rfloor + 1, n-1 < \alpha \leq n, t > a), \quad (2)$$

where ${}^C D$ denotes the Caputo fractional derivative operator, $\lfloor \alpha \rfloor$ is the integer part of $\alpha$. As mentioned above, we use the Caputo definition since its initial conditions have the same form as in the differential equations of integer order [35].

Using the Laplace transform of Mittag-Leffler function, we can get the analytic solution of fractional relaxation-oscillation equations in the following form [36]

$$u(t) = u(0) E_{p,1}(-At^\alpha) + G(t) * f(t); \quad (3)$$

and the analytic solution of oscillation equation can be further expressed as

$$u(t) = u(0) E_{p,1}(-At^\alpha) + u'(0) t E_{p,2}(-At^\alpha) + G(t) * f(t), \quad (4)$$

where $G(t) = t^{p-1} E_{p,p}(-Bt^p)$, and $E_{\alpha,\beta}(Z)$ is the Mittag-Leffler function.

## 2.2. The predictor-corrector approach

The predictor-corrector approach is firstly proposed by Diethelm [18] to solve fractional relaxation-oscillation equations and has since been used in solving various kinds of FDEs. According to its analytical property, the fractional differential equation subject to the initial conditions

is equivalent to the Volterra integral equation

$$\begin{cases} {}_0^C D_t^\alpha y(t) = f(t, y(t)) \\ y^{(k)}(0) = y_0^{(k)}, (k = 0, 1 \cdots, n-1) \end{cases} \quad (5)$$

is equivalent to the Volterra integral equation

$$y(t) = \sum_{k=0}^{\lceil \alpha \rceil - 1} y_0^{(k)} \frac{t^k}{k!} + \frac{1}{\Gamma(\alpha)} \int_0^t (t - \tau)^{\alpha - 1} \cdot f(\tau, y(\tau)) d\tau . \quad (6)$$

We deduce the following formulas under the assumption of a uniform grid $t_n = nh$, where $n = 0, 1, \ldots N$. It is noted that the grid step of $h$, the total duration $T$ and the total grid number $N$ are related by $h = T/N$. The predicted condition $y^p(t_{k+1})$ is obtained by the rectangular rule

$$\int_0^{t_{k+1}} (t_{k+1} - \tau)^{\alpha - 1} f(\tau_j, y(\tau_j)) d\tau \approx \sum_{j=0}^{k} b_{j,k+1} f(\tau_j, y(\tau_j)) , \quad (7)$$

where

$$b_{j,k+1} = \frac{h^\alpha}{\alpha} \left[ (k+1-j)^\alpha - (k-j)^\alpha \right]. \quad (8)$$

And then, we obtain

$$y^p(t_{k+1}) = \sum_{j=0}^{\lceil \alpha \rceil - 1} \frac{t_{k+1}^j}{j!} y_0^{(j)} + \frac{1}{\Gamma(\alpha)} \sum_{j=0}^{k} b_{j,k+1} f(\tau_j, y(\tau_j)) . \quad (9)$$

Applying the trapezoidal quadrature formula, the integration on the right side of Eq. (6) can be replaced by

$$\int_0^{t_{k+1}} (t_{k+1} - \tau)^{\alpha - 1} f(\tau_j, y(\tau_j)) d\tau \approx \sum_{j=0}^{k+1} a_{j,k+1} f(\tau_j, y(\tau_j)), \quad (10)$$

where

$$a_{j,k+1} = \frac{h^\alpha}{\alpha(\alpha+1)} \cdot \begin{cases} (k^{\alpha+1} - (k-\alpha)(k+1)^\alpha), & j = 0 \\ ((k-j+2)^{\alpha+1} + (k-j)^{\alpha+1} - 2(k-j+1)^{\alpha+1}), 1 \leq j \leq k \\ 1, & j = k+1 \end{cases} \quad (11)$$

Using Eqs. (6) ~ (11), one gets

$$y(t_{k+1}) = \sum_{j=0}^{\lceil\alpha\rceil-1} \frac{t_{k+1}^j}{j!} y_0^{(j)} + \frac{1}{\Gamma(\alpha)} (\sum_{j=0}^{k} a_{j,k+1} f(\tau_j, y(\tau_j)) + a_{k+1,k+1} f(\tau_{k+1}, y^p(\tau_{k+1}))). \quad (12)$$

The accuracy of this algorithm is $O(h^q)$ [18], where $q = \min(2, 1+\alpha)$ and $h$ denotes the time step.

In the case of fractional relaxation-oscillation equations, the right-hand side of Eq. (5) is restated as

$$f(t, y(t)) = f(t) - Ay(t). \quad (13)$$

Substituting equation (13) into equation (12) yields

$$y(t_{k+1}) = \sum_{m=0}^{\lceil\alpha\rceil-1} \frac{t_{k+1}^m}{m!} y_0^{(m)} + \frac{1}{\Gamma(\alpha)} (\sum_{m=0}^{k} a_{m,k+1}(f(t_m) - Ay(t_m)) + a_{k+1,k+1}(f(t_{k+1}) - y^p(t_{k+1}))). \quad (14)$$

## 3. FRO toolbox

This section is to introduce the FRO Matlab toolbox we developed in recent years to solve fractional relaxation-oscillation equations, which uses the predictor-corrector approach described in section 2.2. The toolbox is easy-to-use with a friendly graphical user interface for users to choose parameters and to see the results immediately.

To get started, users can enter "FRO" into the Matlab command line. The main interface is presented in Fig. 1 and is divided into two parts consisting of the input and the output modules, respectively.

In the input module, users need to fill in necessary parameters, such as fractional derivative order $\alpha$, relaxation coefficient $A$, time-step $dt$, total duration $t$, initial conditions $y(0)$ and $y'(0)$, and external source $f(t)$. For the sake of convenience, the parameters in the text controls have default values. When clicking the "Reset" button, the parameters will revert to their default values. The default values are listed in Table 1. When clicking "Load" button, the FRO toolbox will plot a curve in terms of input data file "data.xls" which is in the default working path of the FRO file. Users can also distinguish different results using the curve property sub-module to choose distinct linestyles.

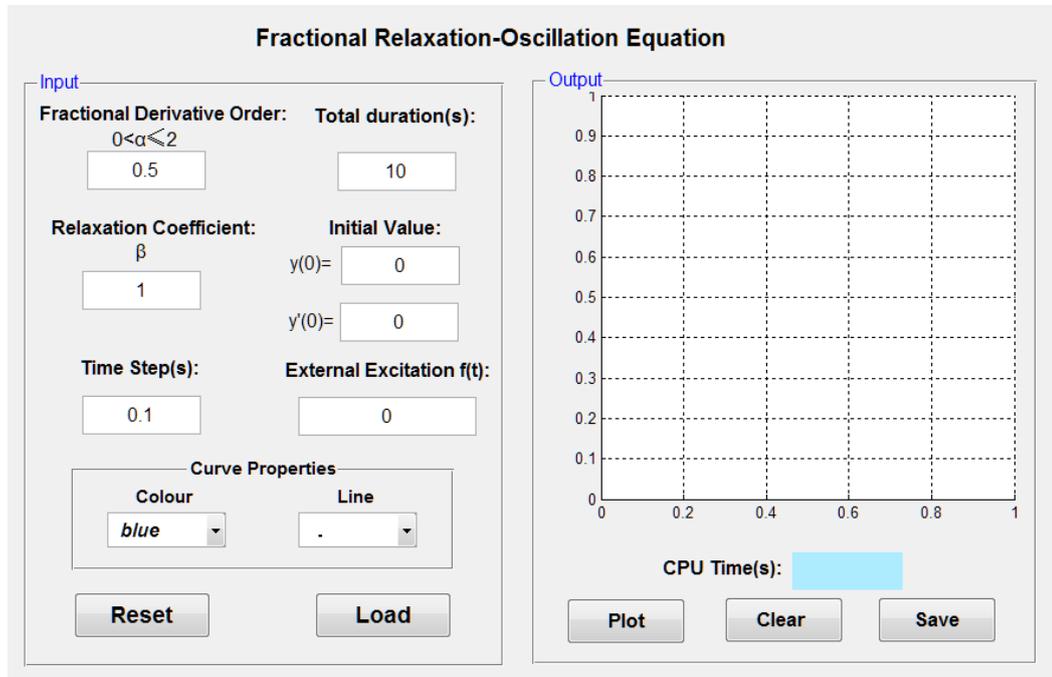

Figure 1. Main user interface

Table 1. Default values of the parameters.

| Parameter | Default Value |
| --- | --- |
| Fractional derivative ($\alpha$) | 0.5 |
| Relaxation coefficient ($A$) | 1 |
| Time-step (dt) | 0.1s |
| Total duration (t) | 10s |
| Initial condition $y(0)$ | 0 |
| Initial condition $y'(0)$ | 0 |
| External function $f(t)$ | 0 |

After inputting parameter of choice into the text controls, users can see the solution curve of the corresponding equation by clicking the "Plot" button. It is noted that the designed range of fractional derivative $\alpha$ is within $(0,2]$. If the input derivative order is outside of this range or disobey the digit format allowed by Matlab, a dialogue as shown in Fig. 2 will pop up. If the existing curves need to be removed, one can click the "Clear" button. For the sake of saving and analyzing data, users can click the "Save" button to preserve the produced figures in a variety of formats, such as *.fig, *.jpg, *.bmp, and *.eps to satisfy varied needs.

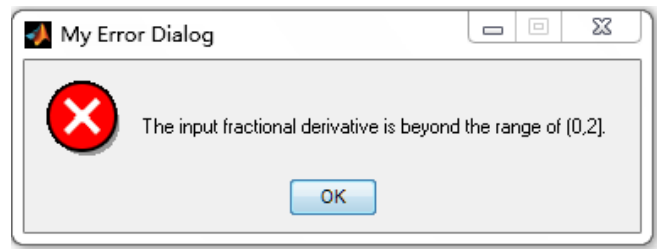

Figure 2. An error reminder.

## 4. Illustration examples

In this section, we demonstrate a few specific examples to illustrate the implementation and use of the FRO toolbox. One user can use this toolbox to get personal experiences and understanding how various parameters of the fractional derivative equation model, such as fractional derivative order, external source function, initial conditions, and relaxation coefficient, influence the relaxation-oscillation behaviors. The following numerical results are all drawn by the FRO toolbox and saved in *.fig format, in which legends are added by Matlab. It should be noted that the units of the parameters need to be uniformed before calculation.

**Example 1:** Eq. (1) with fractional derivative orders $\alpha$ =0.9, 0.8, 0.7 and 0.6, the relaxation coefficient $A=1$, the time step $t$=0.02, the total duration $t$=4s, subject to the initial conditions $y(0)=0$, $y'(0)=0$, the external function $f(t)=5cos(t^2)\cdot exp(-t)$.

Since the FRO toolbox can draw only one curve each time, we start to input the fractional derivative order 0.7 as well as the other parameters into the corresponding text controls. Then we choose a suitable line style via curve property sub-module and click the "Plot" button to get the numerical results. To see the effect of the varied fractional order on the relaxation behavior, we change the fractional derivative order to 0.6, 0.8, 0.9 in the text control while keeping the other parameter unchanged and see the comparable results shown in Fig. 3.

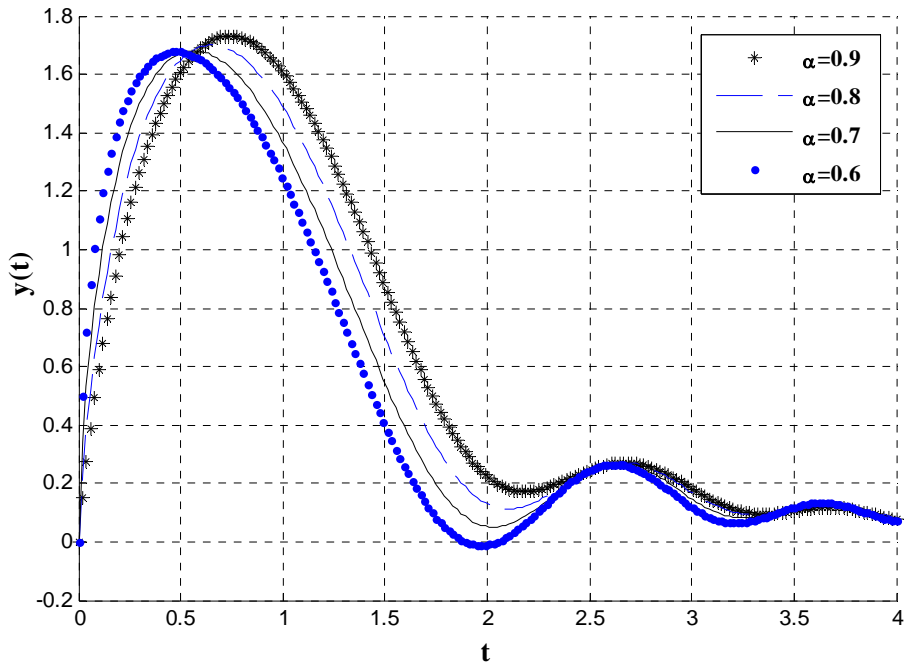

Figure 3. Different fractional relaxation curves with varied fractional orders of 0.9, 0.8, 0.7, and 0.6

One can see from Fig. 3 that the four curves corresponding to four different orders of fractional derivative behave differently with a clear tendency. This toolbox is a very convenient tool for researchers to investigate relaxation behaviors of complex viscoelastic media.

**Example 2:** This illustrative case is to display the influence of initial conditions on oscillation behaviors. Consider the fractional relaxation-oscillation equation (1)

$${}^C_0 D_t^{1.8} u(t) + u(t) = cos(t^2) \cdot exp(-t)$$

with varied initial conditions

(1) $y(0)=1,\ y'(0)=1$;

(2) $y(0)=1,\ y'(0)=0$;

(3) $y(0)=0,\ y'(0)=1$;

(4) $y(0)=0,\ y'(0)=0$.

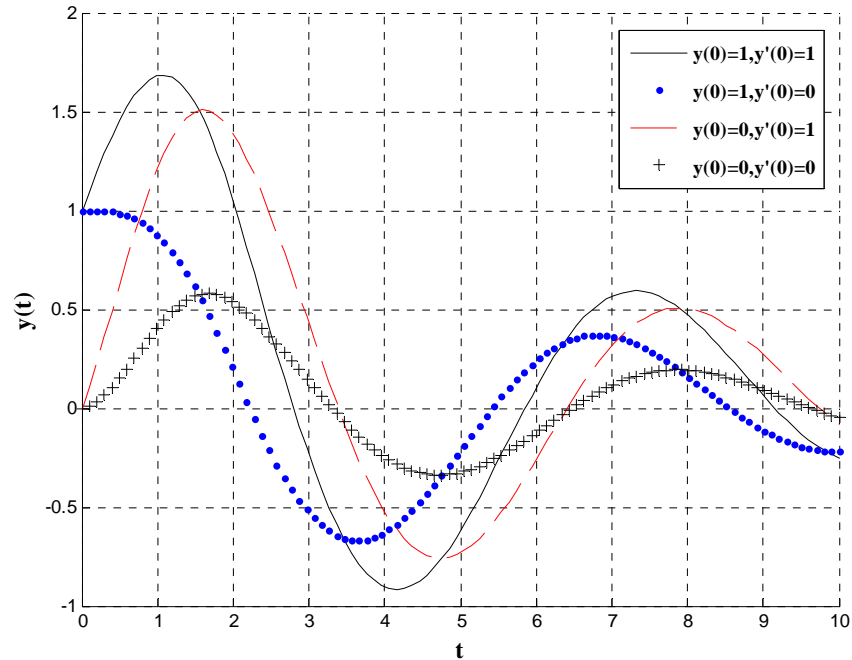

Figure 4. Different fractional oscillation curves with varied initial conditions.

Fig. 4 shows the oscillation curves with four different initial conditions. One can observe that four oscillation curves present the similar damping pattern described by the fractional derivative of order 1.8. In the long term, the influence of initial conditions decays, while that of fractional damping term increases.

**Example 3:** This example is to show how the FRO toolbox can highlight the influence of external force on oscillation behaviors. Consider a governing equation

$${}_{0}^{C}D_{t}^{1.8}u(t)+u(t)=f(t)$$

subject to the initial conditions

$y(0)=1,\ y'(0)=0$,

and the external functions are, respectively, given by

(1) $f(t) = exp(-t)$

(2) $f(t) = \sin(t)$

(3) $f(t) = \cos(t)$

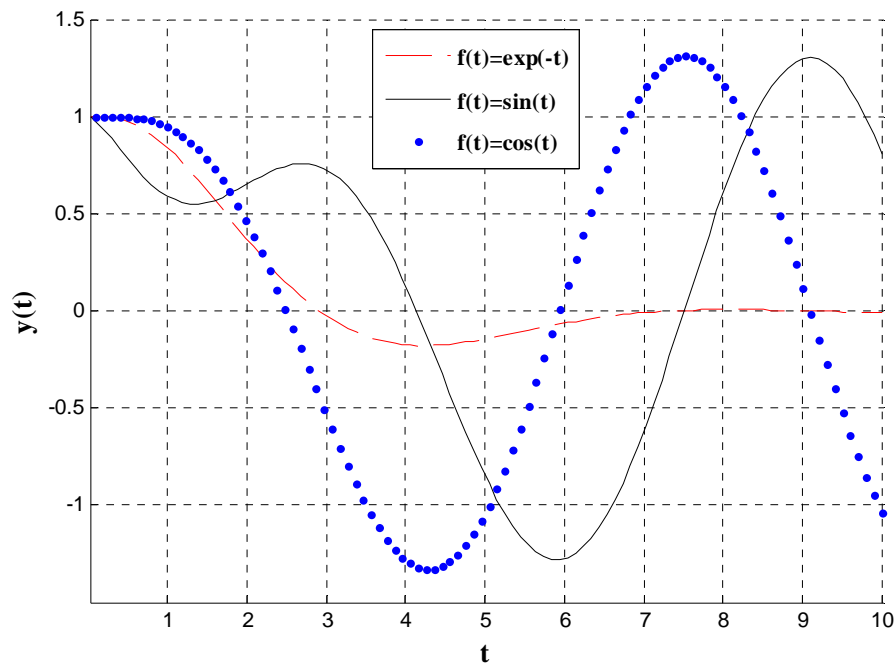

Figure 5. Comparison of fractional oscillation curves with varied external functions.

We can clearly see from Fig. 5 the combined effect of damping and external source term on the resulting oscillation.

**Example 4:** This illustrative case is to show how the FRO toolbox can help practitioners investigate the influence of the relaxation coefficient on relaxation performances. Let us consider the fractional relaxation equation having the fractional order 0.5

$${}_0^C D_t^{0.5} u(t) + Au(t) = t \cdot \sin(t)$$

subject to the initial conditions

$y(0) = 2, \ y'(0) = 0$,

where the relaxation coefficient $A$ varies from 1 to 2 and 3.

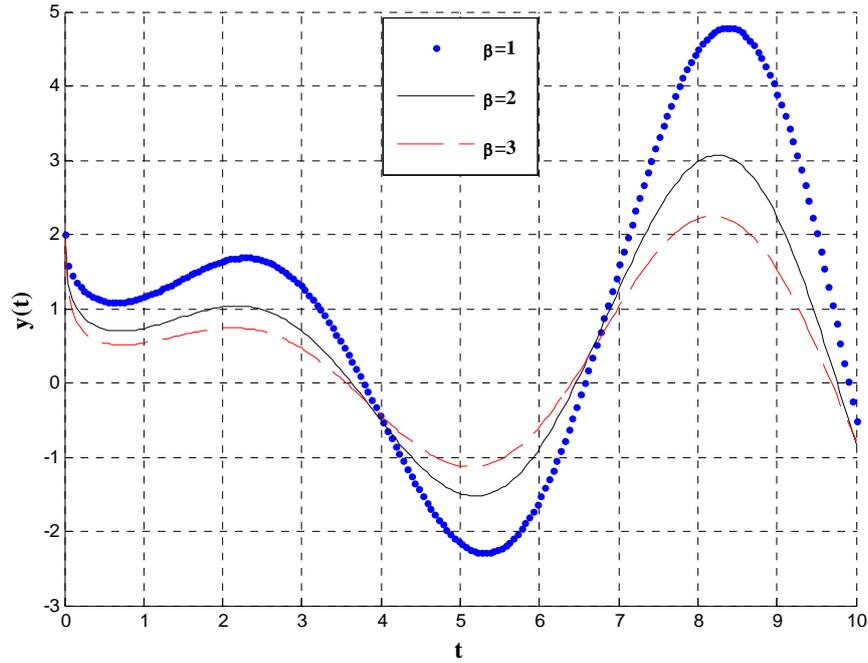

Figure 6. Comparison of fractional oscillation curves with varied relaxation coefficients of 1, 2 and 3.

Fig. 6 shows that the smaller relaxation coefficient, the larger variation of the relaxation curve. These curves are very intuitive to see the influence of relaxation coefficient on the relaxation behaviors.

**Example 5:** This case is to compare numerical results of the FRO toolbox with available experimental data given in ref. [37] for stress relaxation in wheat dough. Let us input the data in Table 2 to produce file "data.xls", and then choose "+" to represent experimental data. Clicking "Load" button, the experimental data curve will be displayed. Then we have relaxation curves of FRO toolbox numerical results by selecting suitable parameters.

Table 2. Stress relaxation data for wheat dough (Cunningham et al, 1953).

| Time(s)    | 0.0   | 1.0   | 2.0   | 4.0   | 6.0   | 8.0   |
|------------|-------|-------|-------|-------|-------|-------|
| Tension(g) | 710.0 | 560.0 | 487.0 | 420.0 | 383.0 | 355.0 |
| Time(s)    | 10.0  | 12.0  | 14.0  | 16.0  | 18.0  | 20.0  |

| Tension(g) | 334.0 | 321.0 | 309.0 | 298.0 | 288.0 | 280.0 |
|---|---|---|---|---|---|---|

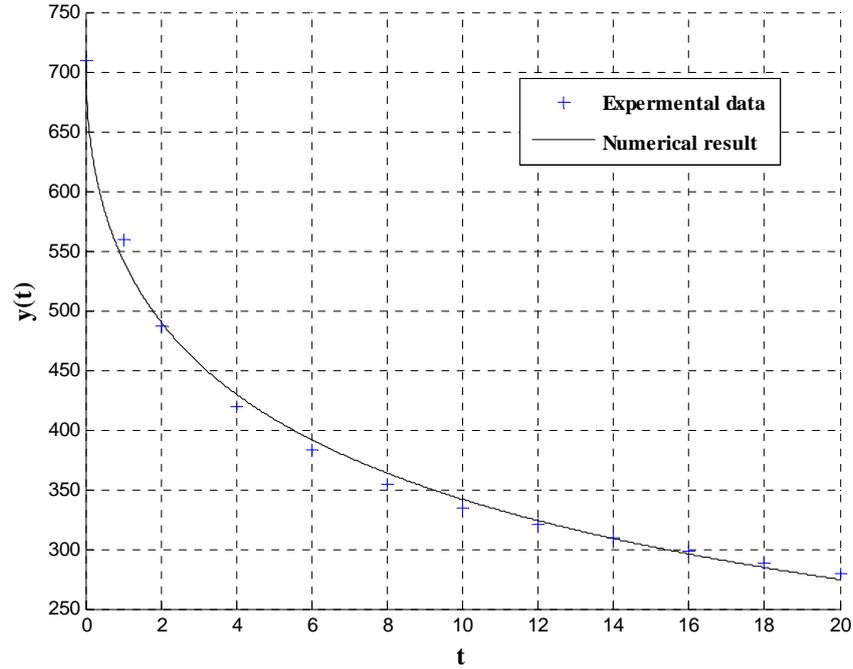

Figure 7. Comparison of numerical result with experimental data (Data from Cunningham et al, 1953).

We can see from Fig. 7 that the numerical curve of the fractional model via the FRO toolbox agrees very well with experimental data. This shows that this toolbox can easily be used in the analysis of lab and field data.

## 5. Remarks

This paper presents the FRO toolbox incorporating useful functions to solve fractional relaxation-oscillation equations. The implementation and use of this toolbox is illustrated in details through five explanatory examples. This FRO toolbox is very easy-to-use with a user-friendly interface for data input and output. The toolbox is very efficient in simulating viscoelastic stress relaxation and damped vibration. With the help of this toolbox, practitioners can easily investigate the performances and behaviors of fractional relaxation-oscillation equation models with varied parameters. This toolbox is of open-source free software and will promote the use and practices of the fractional models in a more broad area of engineering and sciences. The new version of this toolbox is undertaking and includes new functions such as nonlinear source terms to be more applicable to real-world engineering problems.

## Acknowledgement

The authors are grateful of Dr. HongGuang Sun for his valuable help in the preparation of this paper. The work described in this paper was supported by the National Basic Research Program of China (973 Project No.2010CB832702), the National Science Funds for Distinguished Young Scholars (11125208), the Natural Science Foundation of China (11202066), the R&D Special Fund for Public Welfare Industry (Hydrodynamics, Project No. 201101014) and the 111 project under grant B12032.